\documentclass[sort,numbers]{article}

\usepackage{microtype}
\usepackage{graphicx}
\usepackage{subcaption}
\usepackage{booktabs} 
\usepackage[utf8]{inputenc}
\usepackage[T1]{fontenc}
\usepackage{wrapfig}
\usepackage{tcolorbox}
\usepackage{lipsum}
\usepackage{listings}%
\usepackage{textcomp}%
\usepackage{siunitx}%
\usepackage{float}%
\usepackage{tabularx}%

\usepackage{arxiv}
\setcitestyle{authoryear,open={(},close={)}}
\renewcommand{\cite}{\citep}

\usepackage{amsmath}
\usepackage{amssymb}
\usepackage{mathtools}
\usepackage{amsthm}
\usepackage{wrapfig}
\usepackage{pifont}  
\usepackage{xcolor}

\usepackage{tikz}
\usetikzlibrary{arrows.meta,positioning,calc}
\usepackage{multirow}
\usepackage{amsfonts}
\usepackage{mathrsfs}
\usepackage[title]{appendix}
\usepackage{manyfoot}
\usepackage{algorithm}
\usepackage{algorithmicx}
\usepackage{algpseudocode}
\usepackage{array}
\usepackage{makecell}
\usepackage{natbib}
\usepackage[normalem]{ulem}
\usepackage{xurl}
\usepackage{ragged2e}
\usepackage{xltabular}
\usepackage{threeparttable}

\usepackage[capitalize,noabbrev]{cleveref}
\usepackage{comment}
\usepackage{adjustbox}
\usepackage{xspace}
\usepackage{fancyvrb}
\usepackage[subtle]{savetrees}
\usepackage{enumitem}
\usepackage{titling}

\def\shownotes{1}
\ifnum\shownotes=1
\newcommand{\authnote}[2]{[#1: #2]}
\else
\newcommand{\authnote}[2]{}
\fi

\newcolumntype{Y}{>{\raggedright\arraybackslash}X}

\definecolor{CompanionBlue}{HTML}{81A4CD}
\definecolor{CompanionPink}{HTML}{E0B1CB}

\begin{document}
\bibliographystyle{plainnat}

\title{AI Deactivation Reduces Loneliness and Increases Social Interaction}

\author{%
    \large{
        Dora Zhao$^{1,*}$,
        Grace Wang$^{1}$,
        Yutong Zhang$^{1}$,
        Lujain Ibrahim$^{2}$,
        Robert E. Kraut$^{3}$,
        Diyi Yang$^{1,*}$
    }\\[4pt]
    \normalsize{
        $^{1}$Stanford University \quad
        $^{2}$University of Oxford \quad
        $^{3}$Carnegie Mellon University
    }\\[6pt]
    \normalsize{
        $^{*}$Correspondence to:
        \texttt{dorothyz@stanford.edu},
        \texttt{diyiy@stanford.edu}
    }
}

\date{}
\raggedbottom
\maketitle

\begin{abstract}
General-purpose artificial intelligence (AI) assistants are increasingly embedded in everyday life, transforming how people accomplish tasks and potentially reshaping how they interact with others. Prior research has focused on how using AI for explicitly social purposes affects users' psychological well-being and social lives. Yet such uses account for only a small portion of interactions with general-purpose AI assistants. Whether using AI for everyday, instrumental tasks affects users' psychological well-being and social behavior remains unclear. Here we investigate these effects through a longitudinal deactivation experiment with active ChatGPT users (N=625), in which participants were randomly assigned to abstain from using AI assistants for either fourteen days (treatment) or one day (control). Participants used AI predominantly for instrumental purposes, with 84.4\% reporting productivity as their most frequent use case. Two weeks of deactivation significantly reduced loneliness relative to the control group ($\beta = -0.10$, $p=0.03$), with reductions persisting two weeks after participants resumed using AI ($\beta = -0.16$, $p=0.02$). Deactivation also significantly increased the number of people participants interacted with in person and the time they spent socializing over the intervention period. Open-ended responses indicate that during the deactivation period participants sought help from others for tasks they otherwise would have delegated to AI and redirected attention toward social interaction. Overall, our findings suggest that even when AI is used primarily to accomplish instrumental tasks, it may inadvertently reduce opportunities for social interaction and connection.
\end{abstract}

\section{Introduction}
The promise of general-purpose AI assistants is that they can fulfill an ever-expanding range of human needs . Designed as ``universal assistants,'' these systems can answer factual questions, help with complex work tasks, and even provide emotional or social support~\cite{chatterji2025people,anthropic2025affective,tamkin2024clio}. As a result of this versatility, AI assistants have become deeply embedded in many domains of everyday life, changing how people work, seek out information, and form relationships~\cite{zhang2026interaction,burtch2024consequences,cui2026effects,noy2023experimental}. Many of the functions AI assistants now perform were previously fulfilled by other people or completed alongside them. Thus, as AI increasingly takes on these functions, they may also alter how people allocate their time and engage with one another.

Despite this potential, the consequences of general-purpose AI assistance on users' social lives remain poorly understood. Existing work has largely focused on how using AI for explicitly social purposes, such as for companionship or personal advice, affects well-being and social behaviors~\cite{zhang2026interaction,de2026ai,maples2024loneliness,folk2026does,ibrahim2026sycophantic, kirk2025neural}. However, social interactions represent only a small fraction of general-purpose AI use~\cite{anthropic2025affective,phang2025investigating}. For example, recent research categorizing over 1.1 million ChatGPT conversations suggests that the heaviest uses are for practical guidance (28.8\%), seeking information (24.4\%), and writing (23.9\%)~\cite{chatterji2025people} whereas discussion of relationships or personal reflection occurs in only $1.9\%$ of messages.


This predominantly instrumental use of AI assistants may shape human interaction in at least two ways. One possibility is that using AI makes users more efficient, speeding up existing work or automating it altogether, thereby freeing up time that people can redistribute toward socializing or other discretionary activities~\cite{nie2002impact}. Another possibility is that using AI instead displaces interactions with other people, either directly by replacing someone users would otherwise have asked for help, or incidentally, by taking over tasks, such as work or information-seeking, that would otherwise have involved interacting with others. Even seemingly inconsequential everyday interactions contribute to feelings of belonging and happiness~\cite{sandstrom2014efficiency,gunaydin2021minimal}, and losing opportunities for human interaction may increase loneliness, with potential consequences for mental and physical health~\cite{mushtaq2014relationship,hawkley2010loneliness}. The extent of this displacement may depend on how long and how intensively individuals use AI assistants, as sustained reliance could alter their patterns of seeking assistance from others.

In order to understand how AI use shapes the social lives of existing users, we must examine what happens when this technology is no longer available. However, observational studies cannot disentangle AI's effects from preexisting differences in users' social relationships and behaviors~\cite{zhang2026interaction,folk2026does}. Experimental interventions that introduce users to AI can only study the effects for new adopters~\cite{fang2025ai}. Thus, deactivating users --- or temporarily removing access to AI --- provides an opportunity to causally examine how users' lives change in its absence.

\begin{figure}
    \centering
    \includegraphics[width=\linewidth]{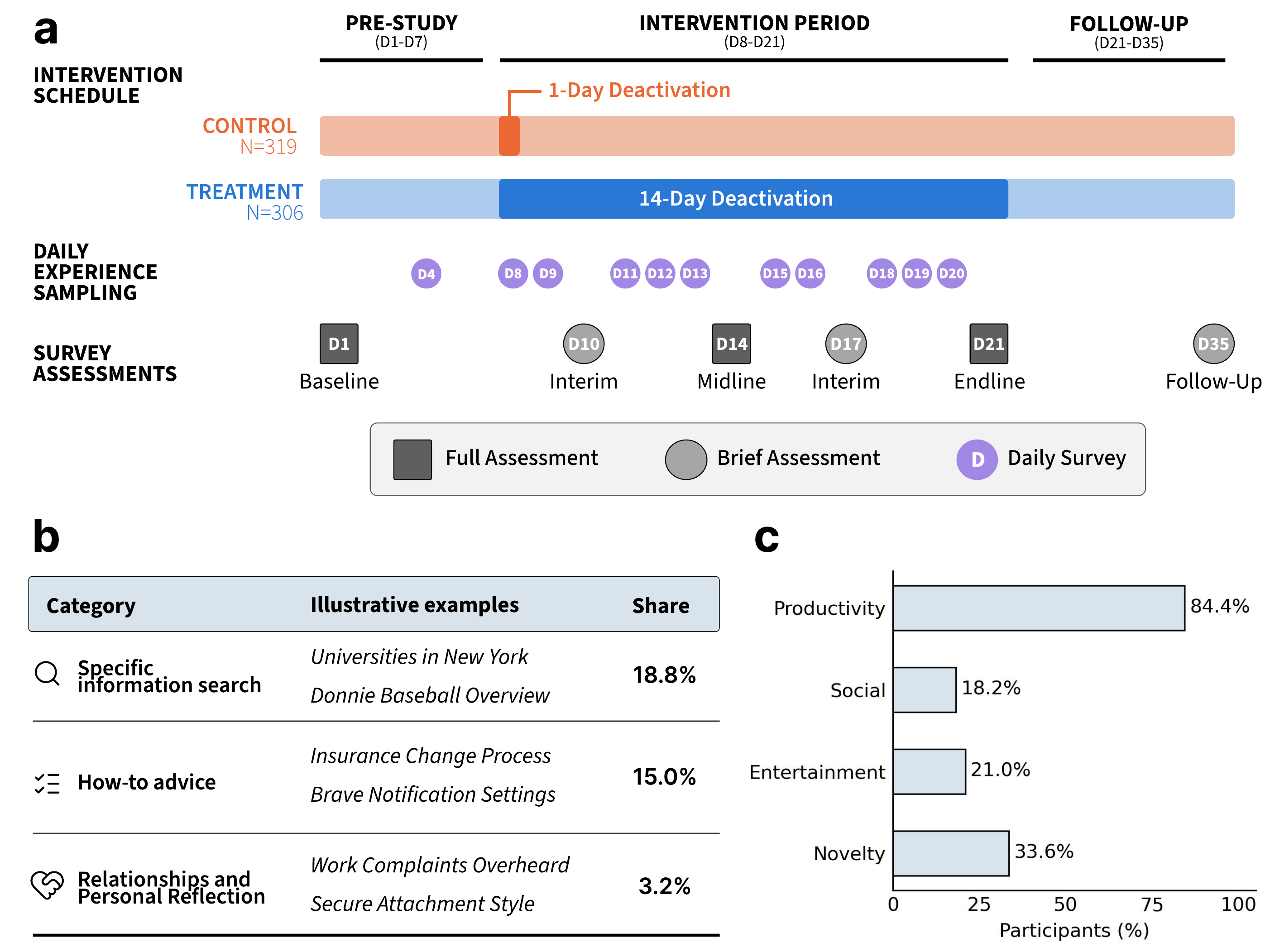}
    \caption{\textbf{Study design and baseline chatbot usage characteristics.} Panel (a) provides an overview of the longitudinal deactivation experiment. Participants were assigned to either a control group, which abstained from AI use for one day, or a treatment group, which abstained for 14 days. Participants completed five survey assessments spanning baseline (Day 1) to endline (Day 21), followed by a post-intervention survey (Day 35). Full assessments were administered at baseline, midline, and endline, with brief assessments at two interim time points and follow-up. Daily experience-sampling surveys were administered once before the intervention began and on intervention days without scheduled survey assessments. Panel (B) shows examples of topics from participants' conversations with AI assistants and the percentage of topics that fall under the category. Panel (C) shows participants' primary purposes of AI use prior to the intervention (productivity, social interaction, entertainment, and novelty). Participants rated the frequency of each purpose independently on a seven-point scale. Bars indicate the percentage of participants for whom each purpose received their highest frequency rating. Ties were allowed, so percentages may sum to more than 100\%.}
    \label{fig:pull}
\end{figure}

To investigate how everyday AI usage shapes social outcomes---beyond cases in which users explicitly seek social connection---we conducted a preregistered, randomized deactivation experiment with active ChatGPT users ($N=625$) in which participants abstained from using AI assistants for an assigned period of time. Our participants primarily used AI for productivity purposes with $84.7\%$ reporting this as their primary use case. Participants were assigned to deactivate for either one day (control) or two weeks (treatment). Deactivation was enforced through our study infrastructure which blocked access to over thirty popular AI assistants. Throughout the two-week period, we collected repeated measures of psychological well-being, daily activities, and social interactions from participants in both conditions through five survey waves and daily experience sampling, alongside direct measures of their AI usage (see Fig.~\ref{fig:pull}). 

We found that two weeks of AI deactivation significantly reduced loneliness relative to the control condition. Deactivation also changed participants' social behavior. Treatment participants reported interacting with more people in person and feeling a greater sense of social support relative to that of the control group over the intervention period. The reduction in loneliness persisted after the intervention concluded; treatment participants reported feeling significantly less lonely than those in the control in a follow-up survey two weeks after the study. These findings are consistent with the possibility that changes in participants' social lives contribute to the reductions in loneliness observed following deactivation. We also observed greater reductions in loneliness for participants who had used ChatGPT for a longer period of time but no relationship between daily usage and changes in loneliness. Finally, despite these changes in social interaction, we found no significant effect of deactivation on total time spent working, suggesting that increased social engagement did not come at the expense of work.

Overall, our findings suggest that general-purpose AI assistants may displace opportunities for human connection, even when used for instrumental rather than relational purposes. As AI assistants become more ubiquitous in everyday life, understanding their broader social effects requires examining how sustained everyday use shapes people's lives. Randomized deactivation experiments provide an approach for causally estimating these effects among existing users. More broadly, our findings underscore the importance of designing these technologies with attention to how they shape opportunities for human connection and collaboration, rather than solely to their ability to complete individual tasks.

\section{Results}
We conducted a two-week randomized deactivation experiment with 625 active U.S.-based ChatGPT users. To be eligible for the study, participants had to meet the following criteria: (1)~use ChatGPT for at least 15 minutes per day; (2)~have maintained an account on the platform for at least one month; (3)~use Chrome as their primary browser; and (4)~be willing to accept a payment of \$40 per week to abstain from using AI. Following prior deactivation studies~\cite{allcott2025effect}, we assigned participants to abstain from using AI for either one day (control) or two weeks (treatment). The one-day control was intended to reduce demand effects and mitigate differential attrition. In addition to our baseline assessment, we collected data through four survey waves and daily experience-sampling assessments from all participants during the two-week study period. Two weeks after the intervention ended, we conducted a follow-up survey. 

Our participants predominantly used AI for instrumental or productivity-related purposes. $84.4\%$ of participants reported using ChatGPT most frequently for productivity-related purposes, and $52.8\%$ reported using ChatGPT for productivity purposes at least once per day. Participants' conversation history also reflected this pattern; the most common use categories included practical guidance (e.g., advice-seeking) and seeking information (e.g., looking for specific information, searching about purchasable products). Over a fifth of participants (N=126, $20.2\%$) reported that they had never used AI for social purposes, and topics related to relationships or personal reflection comprised only $3.2\%$ of the conversations collected. 

\begin{figure}
    \centering
\includegraphics[width=\linewidth]{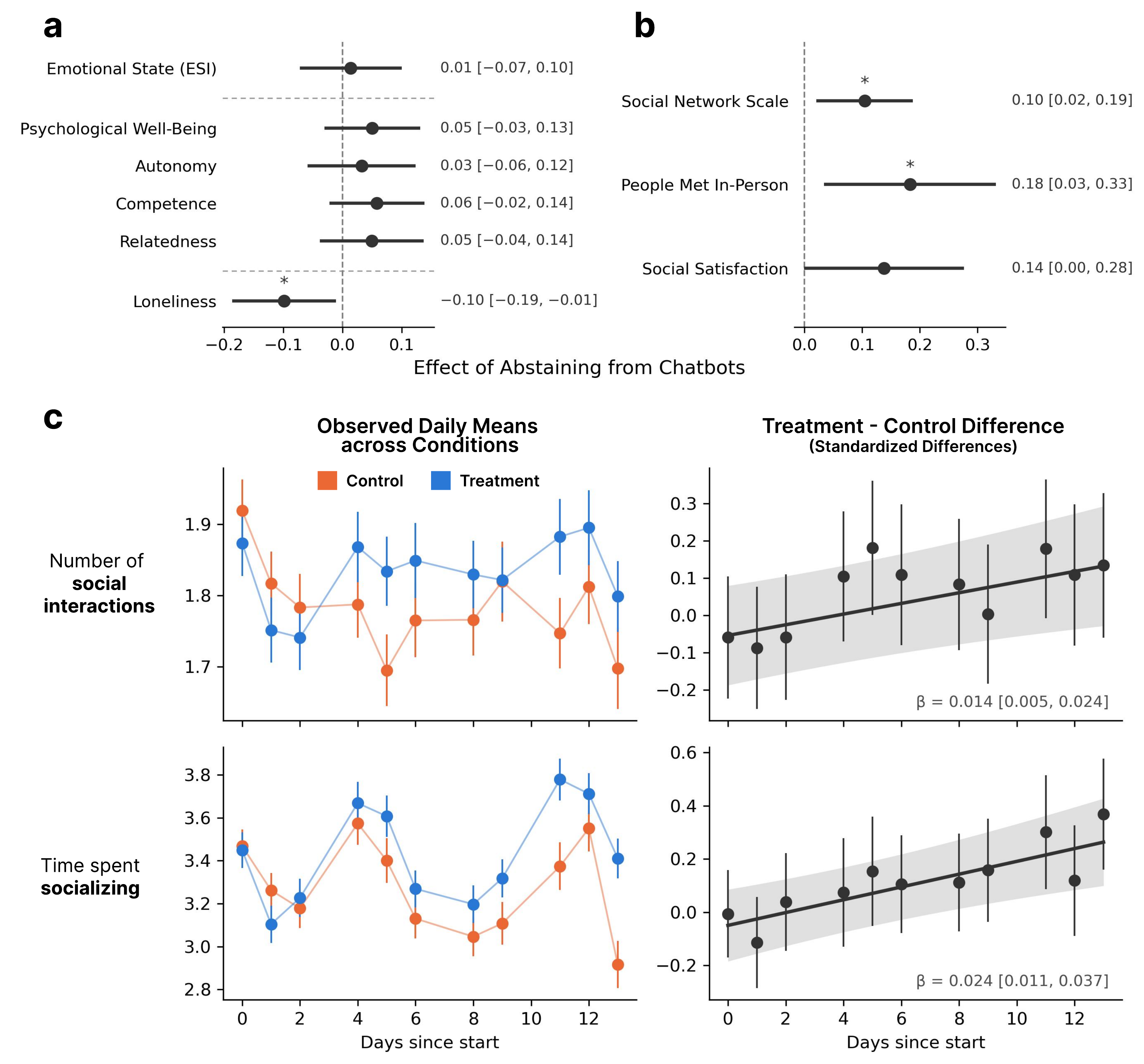}
    \caption{\textbf{Deactivating AI reduces loneliness and increases social interaction.} Estimated treatment effects from linear mixed-effects models comparing changes from baseline to the intervention period between treatment (N = 306) and control (N = 319) participants, adjusting for covariates and including participant-level random intercepts. Panel (a) shows the differences between treatment and control groups in changes from baseline during the intervention period for our primary outcomes --- emotional state, psychological well-being, and loneliness --- estimated using data from five survey waves (baseline, midline, endline, and two interim surveys). Panel (b) visualizes the differences between conditions for our secondary social outcomes, estimated using data from three survey waves (baseline, midline, and endline). For both Panels (a) and (b), points represent treatment effect estimates, horizontal bars indicate 95\% confidence intervals, and asterisks denote statistical significance (* $p < 0.05$, ** $p < 0.01$, *** $p < 0.001$). Panel (c) shows the daily trajectories of social interactions (top) and time spent socializing (bottom), measured through experience sampling. Left panels show mean daily responses by experimental condition. Right panels show daily treatment–control differences (points) and fitted linear trends (solid lines). Error bars and shaded bands represent 95\% confidence intervals.}
    \label{fig:primary}
\end{figure}
\subsection{Deactivating AI leads participants to feel less lonely and increases social interaction}
We first examined how abstaining from AI use affected participants' social and psychological well-being. Outcomes were assessed in the baseline survey pre-intervention and at four points during the two-week intervention period. To estimate treatment effects, we fit linear mixed-effects models with condition, intervention period, and their interaction as fixed effects as well as participant-level random effects. Throughout our analyses, we report the treatment-by-period interaction coefficient as our primary treatment effect estimate, representing the difference between treatment and control groups in their average change from baseline to the intervention period. All reported coefficients ($\beta$) are expressed in baseline standard-deviation units. When reporting results, we used the Benjamini-Hochberg procedure and report adjusted $Q$-values when accounting for multiple hypothesis tests.\footnote{Loneliness and emotional state were preregistered as individual directional hypotheses and are reported without adjustment. Q-values are reported for outcomes within the preregistered psychological well-being family and our secondary outcomes related to social behavior.}

Participants assigned to deactivate for two weeks experienced a significantly greater reduction in loneliness than participants in the control ($\beta=-0.099$, $p=0.028$) (see Fig.~\ref{fig:primary}a). In addition, AI deactivation was also associated with broader changes in participants' social outcomes. As shown in Fig.~\ref{fig:primary}b, participants in the treatment condition reported greater social support ($\beta=0.104$, $p=0.015$, $Q=0.037$) and more frequent in-person interactions ($\beta=0.183$, $p=0.019$, $Q=0.037$). They also experienced greater satisfaction with their social interactions ($\beta=0.137$, $p=0.053$, $Q=0.066$), although this effect is not statistically significant. 

To examine how these changes in social behavior unfolded over the intervention, we analyzed participants' daily experience-sampling reports. Treatment participants reported engaging in more social interactions each day compared to control participants ($\beta=0.153$, $p=0.007$), with this difference widening over the intervention period ($\beta=0.014$, $p=0.004$). Treatment participants also showed a larger increase in time spent socializing from baseline relative to control participants, although this difference was not statistically significant ($\beta=0.130$, $p=0.077$). The difference between the treatment and control groups grew significantly over the course of the intervention ($\beta=0.024$, $p<0.001$). Changes in social behavior, however, were not accompanied by changes in working time. We found no significant difference between treatment and control participants in daily time spent working ($\beta=-0.045$, $p=0.576$), nor in how working time changed over the intervention period ($\beta=-0.005$, $p=0.381$).

Despite these improvements in social well-being, we found no significant effects of abstaining from AI use on participants' emotional state or broader psychological well-being. This pattern is consistent with prior longitudinal studies of AI use over similar timescales, which have also found limited evidence of changes in psychological well-being~\cite{ibrahim2026sycophantic,kirk2025neural}. These results suggest that the effects of deactivation may be concentrated in participants' social lives, rather than extending to broader dimensions of psychological well-being, at least over the two-week intervention period.

\subsection{Changes in social interaction may contribute to declines in loneliness}
\begin{figure}
    \centering
    \includegraphics[width=\linewidth]{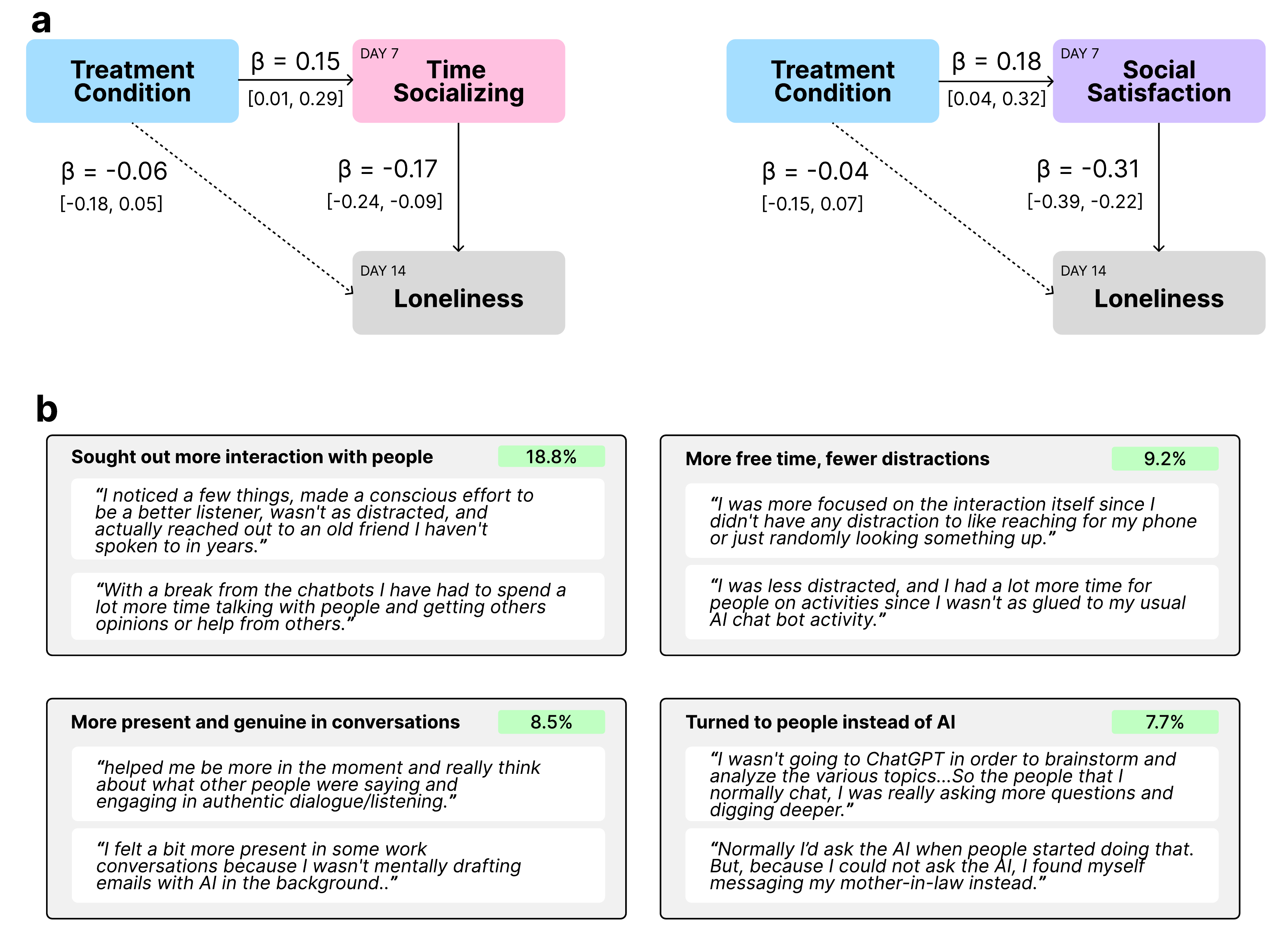}
    \caption{\textbf{Changes in social interaction may help explain reductions in loneliness following AI deactivation}. In Panel (a), we show exploratory longitudinal mediation analyses examining whether changes in social interaction one week into the intervention mediated the effect of AI deactivation on loneliness at the end of the intervention, adjusting for baseline measures. Separate models examined time spent socializing (left) and social satisfaction (right) as potential mediators. Arrows indicate standardized regression coefficients, with 95\% confidence intervals in brackets. Panel (b) shows themes identified from running LLM-based topic-modeling on treatment participants' open-ended reflections on how their social interactions changed during the intervention. Four illustrative themes are shown with response percentages and representative participant quotations.
    }
    \label{fig:mediation}
\end{figure}
Next, to understand whether these changes in participants' social lives contributed to the reductions in loneliness, we conducted exploratory longitudinal mediation analyses. We examined whether the participants' social behaviors one week into the intervention mediated the effect of our intervention on loneliness at the end, adjusting for baseline levels. Specifically, we examined both the \emph{quantity} and \emph{quality} of social interaction. As shown in Fig.~\ref{fig:mediation}a, we found significant indirect effects through both increased time spent socializing  (indirect effect = $-0.025$, 95\% CI [$-0.057$, $-0.001$]) and increased social satisfaction (indirect effect = $-0.054$, 95\% CI [$-0.104$, $-0.010$]). These findings suggest that changes in both the quantity and quality of social interaction one week into the intervention may have contributed to subsequent reductions in loneliness.

To better understand how abstaining from AI use related to these changes in social behaviors, we analyzed treatment participants' open-ended responses from the endline survey reflecting on how they perceived their social interactions to have changed during the two-week intervention period. We used automated topic modeling techniques to cluster their responses. $41.5\%$ of responses discussed improved social interactions while $42.3\%$ reported little or no change and $8.8\%$ reported worse or mixed social interactions. In addition, topics emerged related to how deactivation affected their social lives. In $18.8\%$ of responses, participants described seeking more interaction with others, such as spending more time in conversations they were having or reaching out for additional social interaction (see Fig.~\ref{fig:mediation}b). Furthermore, in $7.7\%$ of responses, participants explicitly mentioned turning to people for assistance or interaction they would have otherwise sought from AI, including asking friends, family members, or colleagues questions that they otherwise would have asked an AI system. In addition to the increased quantity of social interaction, participants also attributed improvements to having more free time or fewer distractions as they were spending less time on their device (9.2\%), or feeling more present and genuine during conversations (8.5\%).

\subsection{Longer-term ChatGPT users experience greater reductions in loneliness \\during deactivation}
We next examined whether the effects of deactivation varied with participants' pre-intervention patterns of AI use. We focused on two dimensions of baseline use: \emph{time spent}, measured as participants' self-reported daily time spent interacting with AI, and \emph{account tenure}, measured as the length of time participants reported having a ChatGPT account. Daily time spent using AI did not moderate the effect of deactivation on loneliness ($\beta=0.002$, $p=0.965$). By contrast, account tenure significantly moderated the effect ($\beta=-0.109$, $p=0.021$). For participants in the treatment condition, those who had used ChatGPT for longer experienced greater reductions in loneliness relative to the control condition. That is, the effects of the intervention were more strongly associated with the duration of participants' prior AI use than with the amount of time they typically spend using AI each day. One possible explanation is that, with longer-term use, AI becomes more deeply integrated into participants' routines, such that abstaining from using it produces more pronounced downstream changes.

\subsection{Effects of AI deactivation persist two weeks after the intervention}
\begin{figure}
    \centering
    \includegraphics[width=0.85\linewidth]{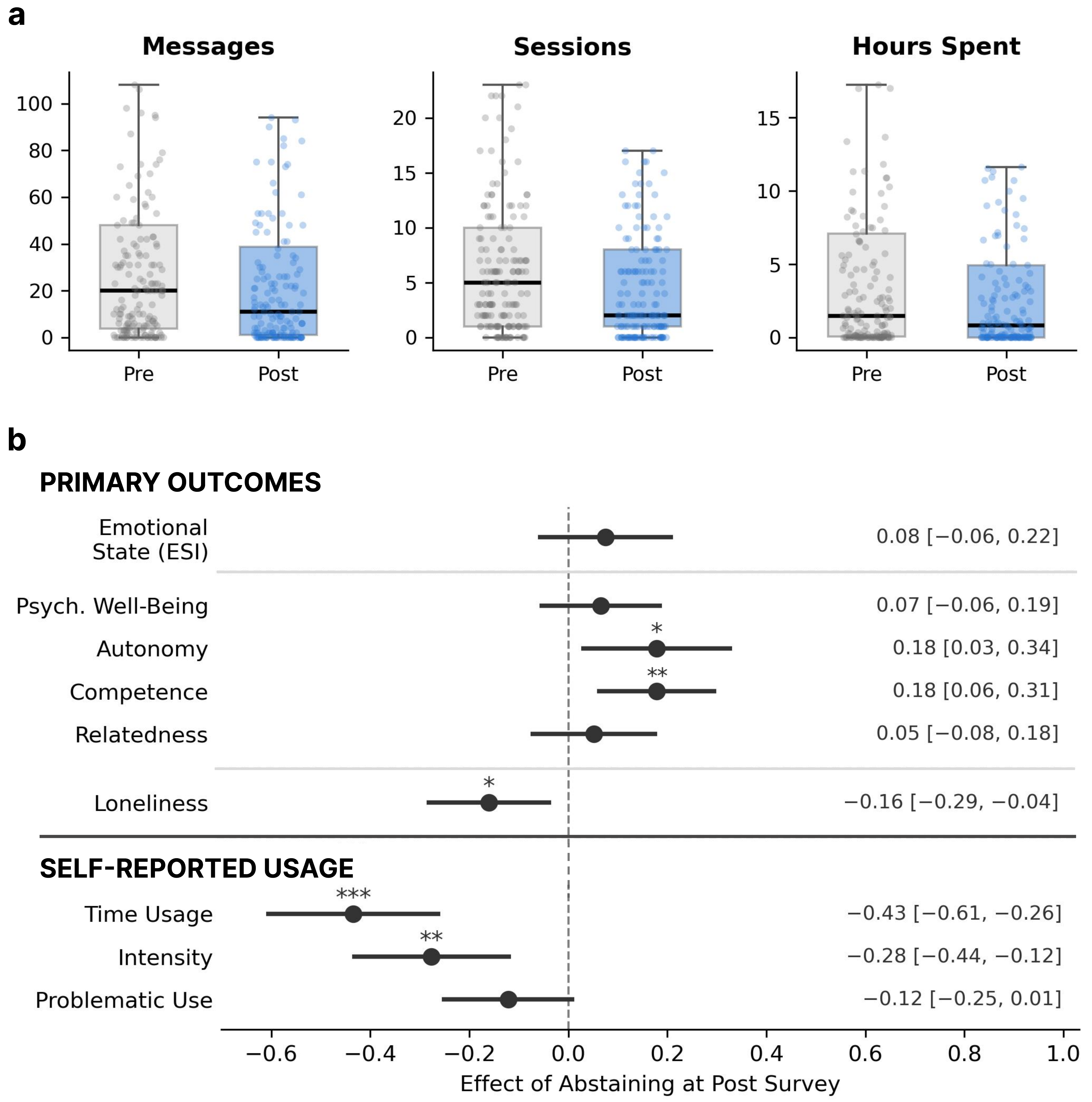}
    \caption{\textbf{Deactivation leads to sustained reductions in AI use and loneliness two weeks after the intervention concludes.} In Panel (a), we visualize usage in the two weeks before (Pre) and two weeks after (Post) the intervention period for treatment participants who shared their usage metadata (N=172). Usage is measured by the number of messages sent, sessions initiated, and total hours spent using ChatGPT. In Panel (b), we report the estimated treatment effects on our primary outcomes and self-reported AI usage at the post-intervention follow-up for participants who completed the post-survey (N=442). Points represent treatment effect estimates, horizontal bars indicate 95\% confidence intervals, and asterisks denote statistical significance (* $p < 0.05$, ** $p < 0.01$, *** $p < 0.001$).}
    \label{fig:placeholder}
\end{figure}
Finally, two weeks after the intervention ended, we conducted a post-study survey to assess whether changes persisted after participants resumed their usual AI use. Participants ($N=442$; Treatment: $N=233$; Control: $N=209$) reported their current AI use, emotional state and well-being using the same measures administered at baseline. Participants could also optionally share their ChatGPT usage metadata ($N=326$; Treatment: $N=172$; Control: $N=154$), providing behavioral measures of time spent on ChatGPT and the number of messages sent. We used an ANCOVA comparing the post-study to pre-intervention results. We report changes in both participants' usage patterns and their psychological well-being at the follow-up.

\paragraph{Reductions in AI usage among treatment participants continue even after the abstinence intervention.} Participants in the treatment condition reported significantly lower time usage ($\beta=-0.434$, $p<0.001$, $Q<0.001$) and interaction intensity  ($\beta=-0.276$, $p=0.001$, $Q=0.001$) compared to those in the control condition even two weeks after the intervention ended. There was no significant difference in the amount of problematic usage between conditions ($\beta=-0.121$, $p=0.079$, $Q=0.079$), although both groups reported lower amounts during the post-study follow-up. 

Behavioral usage data showed directionally similar patterns. Relative to their pre-intervention measurements, participants in the treatment condition sent an average of $3.7\pm4.6$ fewer messages, initiated $1.2\pm0.4$ fewer chat sessions, and spent $67.6\pm43.1$ fewer minutes on ChatGPT during the two-week follow-up period.\footnote{Descriptive statistics are reported in their original units, whereas inferential analyses were conducted on log-transformed usage measures.} However, these changes did not significantly differ between participants in the treatment and control conditions (messages: $\beta=-0.036$, $p=0.855$, $Q=0.855$; time spent chatting: $\beta=-0.114$, $p=0.800$, $Q=0.855$; sessions: $\beta=-0.038$, $p=0.686$, $Q=0.451$). Overall, these findings indicate a persistent shift in participants' perceptions of AI use, with behavioral measures showing directionally consistent, but statistically nonsignificant, reductions in use.

\paragraph{Effects on loneliness persist after two weeks and changes in psychological needs emerge.} Turning to downstream effects on participants' well-being, the reduction in loneliness persisted two weeks after the intervention ended. Adjusting for baseline levels, participants in the treatment condition reported significantly lower loneliness than those in the control condition ($\beta=-0.164$, $p=0.016$, $Q=0.047$). They also reported greater feelings of competence ($\beta=0.181$, $p=0.004$, $Q=0.022$) and autonomy ($\beta=0.181$, $p=0.025$, $Q=0.050$). Notably, differences in competence and autonomy were not significant at the end of the intervention but emerged at the two-week follow-up. This difference could suggest that some effects of the intervention on psychological needs may develop over a longer timescale and are not directly observable at the end of the two-week period. Similar to the end of the intervention period, we still observed no significant differences between conditions in emotional state ($\beta=0.076$, $p=0.294$) or other measures of psychological well-being ($\beta=0.065$, $p=0.320$).

\section{Discussion}
Prior research on AI and social well-being has largely focused on explicitly relational uses, such as AI companionship, where AI may directly substitute for human relationships. Yet general-purpose AI assistants are primarily used for instrumental tasks, raising the question of whether they might also shape users' social lives in less obvious ways. Here we conducted a longitudinal randomized deactivation experiment with participants who primarily used AI for instrumental purposes. We found that deactivating AI usage increased social interaction and reduced loneliness, suggesting that AI use may inadvertently displace opportunities for social interaction, even when that use is not explicitly social.

What might explain these changes in loneliness and social behavior? One possible explanation is social displacement, where time spent using technologies can replace opportunities for interpersonal interaction~\cite{kraut1998internet,nie2001sociability}. Prior social media deactivation studies provide evidence consistent with this theory, finding that abstaining from social media can increase offline social activities and improve subjective well-being~\cite{allcott2020welfare,allcott2026effect}. Beyond displacing time that might otherwise be spent socializing, AI assistants may also displace interpersonal interaction by fulfilling the instrumental assistance people would otherwise seek from others. Tasks such as deciding what product to buy or brainstorming ideas for work are not inherently social, but they can prompt opportunities for interpersonal exchange. Such exchanges can provide benefits beyond practical assistance, including social connection and support~\cite{cohen1985stress}, and even minimal social interactions can improve affect and feelings of belonging~\cite{sandstrom2014efficiency,gunaydin2021minimal,ascigil2025minimal}. Even when human support is available, AI assistants provide people with a convenient, always-available alternative that they can turn to without having to incur the interpersonal costs of seeking help~\cite{ta2020user,branley2023chatbots}. In doing so, AI assistants may inadvertently reduce these opportunities for social interaction.

Our findings are consistent with this account. Participants who deactivated for two weeks interacted with more people and spent more time socializing, alongside reductions in loneliness compared to the control condition. In their open-ended reflections on their perceived changes in social interactions, some participants described turning to others for help that they would otherwise have sought from AI assistants, such as asking family members for recipes or brainstorming with colleagues. Exploratory mediation analyses further suggested that changes in socializing and social satisfaction may help account for reductions in loneliness. 

Whereas prior research on digital technologies has examined whether online interactions displace face-to-face relationships~\cite{kraut2002internet,valkenburg2007online,hall2022social}, and more recent work has considered whether AI companionship substitutes for human relationships~\cite{zhang2026living}, we extend this literature to technologies used primarily for instrumental purposes. Our findings suggest a broader form of displacement that does not require users to seek social connection with AI in the first place. By fulfilling instrumental needs, general-purpose AI assistants may eliminate the accompanying interpersonal exchanges through which social relationships are maintained and strengthened. Although earlier technologies have similarly enabled people to accomplish tasks independently, the breadth of functions supported by general-purpose AI may extend this displacement across an increasingly wide range of activities.

We also found that the reductions in loneliness were greater among longer-tenured users, suggesting that this relationship may evolve as these technologies become more deeply integrated into everyday life. This highlights the theoretical importance of studying \textit{established} AI use rather than focusing solely on initial adoption~\cite{fang2025ai,phang2025investigating}. In addition, as the population of non-AI users shrinks, it will become increasingly difficult to experimentally investigate the effects of AI through initial adoption. By temporarily withholding access from existing users, deactivation experiments offer an important complementary approach for causal inference that may become increasingly necessary as AI assistants become embedded in everyday life and influence users' well-being, time allocation, and offline social activity~\cite{allcott2020welfare,castelo2025blocking,mosquera2020economic}.

Overall, our findings have broader implications for how we design general-purpose AI assistants. Although we employed temporary deactivation to investigate the effects of AI use, complete abstention is unlikely to be a practical or desirable intervention, given the benefits AI assistance can provide~\cite{noy2023experimental,brynjolfsson2025generative}. Instead, our findings underscore the importance of designing AI assistants that proactively support social interaction. As AI becomes more capable of automating everyday tasks, optimizing primarily for individual task performance may overlook the value of human interactions that AI replaces. Future work should continue to investigate how AI can facilitate collaboration between people, such as by helping find common ground in conversations~\cite{tessler2024ai,argyle2023leveraging} or improving the efficacy of human teams or support~\cite{dell2025cybernetic,sharma2023human}. Ultimately, the success of general-purpose AI assistants should be evaluated not only by what they enable individuals to accomplish, but also by how they shape opportunities for people to connect and collaborate with one another.

\section{Methods}
\subsection{Study Design} 
We conducted a randomized experiment to examine the effects of abstaining from AI assistant use on participants' psychological well-being and social behaviors. Participants were randomly assigned to either a treatment condition, in which they were asked to abstain from using AI assistants for 14 days, or a control condition, in which they were asked to abstain for one day. The study was conducted over two waves, from May 6, 2026 -- June 9, 2026 (N=105) and from June 2, 2026 -- July 7, 2026 (N=520). The study protocol was approved by the Stanford University Institutional Review Board and preregistered on the \href{https://osf.io/gaqyb/overview?view_only=9f83cb9db25d4102965844cc98c007ef}{Open Science Foundation (OSF)}.\footnote{We report the following deviations from our pre-registration: (i)~The pre-registration specified baseline-adjusted intent-to-treat (ITT) ANCOVA models comparing outcomes at baseline and endline. For our main analyses, we used a longitudinal linear mixed-effects model incorporating all available measurement waves. To assess the robustness of our findings to this modeling choice, we additionally report inverse-probability-weighted (IPW) ANCOVA estimates of endline treatment effects, accounting for differential attrition (Supplementary Information, Section 4). The resulting estimates were similar in magnitude and direction to those of our primary analyses; (ii)~we included baseline demographic characteristics in our primary models to account for participant heterogeneity and improve precision. Results remained consistent when excluding these additional covariates (Supplementary Information, Section 4); (iii) we report local average treatment effect analyses in Supplementary Information, Section 4. 
}

\paragraph{Participant Recruitment and Selection.} 
We screened 7,900 U.S.-based ChatGPT users via the crowd-sourcing platform Prolific. To qualify for the study, participants had to meet the following criteria: (1)~self-report as using ChatGPT as their primary AI assistant; (2)~use ChatGPT for at least 15 minutes per day; (3)~have used ChatGPT for at least one month; (4)~use Chrome as their primary browser. Finally, these participants (N=2,895) were also willing to accept a payment of \$40 per week to abstain from using AI. Of those screened, 1,250 participants met the screening criteria and were invited to complete the pre-study survey. In total, 625 participants were included in the study, after excluding five participants who failed our attention checks. These participants were then block-randomized into the treatment and control conditions based on their intensity of usage and account tenure.

\paragraph{Study Infrastructure.}
Our study infrastructure, a Chrome browser extension, was used to both deliver our intervention by blocking access to AI assistant services and to track compliance. We compiled 36 URLs of common AI assistant services (see Supplementary Information, Section 1, for full list). The browser extension blocked access to each listed domain and all URLs under that domain, including subpaths (e.g., both chatgpt.com and chatgpt.com/abc). To detect violations, we inspected the ChatGPT API response, which lists the most recent conversations that the user has. These include conversations the user had on their account across all devices. If the user sent a message during their assigned deactivation period, we noted the violation, recording the earliest date of violation.

\paragraph{Survey Design and Measures.}
Participants completed three main surveys: a baseline survey before the intervention (Day 0), a midline survey one week into the intervention period (Day 14), and an endline survey at the conclusion of the intervention period (Day 21). We also collected two shorter interim surveys on Days 10 and 17.

The baseline survey collected demographic information, prior AI usage patterns, and baseline measures of all study outcomes. Our primary outcomes captured psychological and social well-being: loneliness~\cite{russell1996ucla}, emotional state~\cite{allcott2020welfare}, subjective well-being~\cite{su2014development}, and basic psychological need satisfaction and frustration~\cite{chen2015basic}. Secondary outcomes included social behaviors (social support~\cite{lubben2006performance}, social satisfaction, and frequency of social interactions), work activities (time spent working and types of work performed), and AI usage patterns. The midline and endline surveys repeated both primary and secondary outcome measures and included brief open-ended reflection questions, while the interim surveys collected only primary outcomes.

Participants also completed eleven daily experience-sampling surveys reporting their emotional state, number of people with whom they interacted, time spent socializing, and time spent working. Finally, two weeks after the endline survey, we invited participants to complete a follow-up survey assessing the persistence of changes in primary outcomes and AI usage behavior (number of messages sent, number of sessions, time spent on platform).

\subsection{Attrition}
Of the 625 participants who completed the baseline survey and were randomized, 500 (80.0\%) completed the endline survey, corresponding to an overall attrition rate of 20.0\%. 263 participants in the treatment condition (85.9\%) and 237 participants in the control condition (74.3\%) completed the endline survey, a difference of 11.7 percentage points. Of these 500 participants, 433 (86.6\%) also completed the follow-up survey administered two weeks after the intervention. We used full-information maximum likelihood (FIML) under a missing-at-random (MAR) assumption to account for incomplete observations. Full attrition analyses are reported in Supplementary Information, Section 3. 

\subsection{Compliance}
We assessed deactivation compliance using behavioral monitoring and participants' self-reports. Our browser extension monitored participants' ChatGPT activity to detect messages sent during their assigned deactivation period. We classified participants as non-compliant if a violation was detected or they self-reported using AI assistants during their assigned deactivation period. Among the 500 participants who completed the endline, 431 (86.20\%) were fully compliant. Compliance was lower in the treatment condition, where 214 of the 263 participants who completed the endline fully complied (81.36\%). 217 of 237 control participants (91.56\%) completed the full one-day deactivation. 

\subsection{Participants and Procedures}
Participants provided informed consent at the beginning of the study. In total, participants were paid \$22.25 for completing all of the surveys in the study. Treatment participants received a bonus of \$80 if they complied with the deactivation; control participants received \$5.71 (prorated at the \$40 per week rate). We recruited a sample size of 625 participants to detect an effect size of at least $0.3$ with $\alpha=0.05$, assuming a $20\%$ attrition rate.

\bibliography{sn-bibliography}

\paragraph{Author Contributions.} D.Z.: Conceptualization, Methodology, Data Curation, Software, Investigation, Formal Analysis, Writing - original draft, Writing - Review \& Editing. G.W.: Methodology, Data Curation, Investigation, Writing - Review \& Editing. Y.Z.: Conceptualization, Methodology, Data Curation, Writing - Review \& Editing. L.I.: Writing - original draft, Writing - Review \& Editing. R.K.: Methodology, Supervision, Writing - Review \& Editing. D.Y.: Conceptualization, Methodology, Supervision, Funding Acquisition, Writing - Review \& Editing.

\paragraph{Acknowledgments.} D.Z.'s PhD is supported by the Paul \& Daisy Soros Fellowship for New Americans and the Stanford Interdisciplinary Graduate Fellowship. We thank Meryl Ye, Jeffrey Hancock, Hunt Allcott, Carol Dweck, Ashish Mehta, and Michael Bernstein for helpful conversations. This work received funding support from the Stanford Institute for Human-Centered AI (HAI), the OpenAI Mental Health Grant, and NSF grant 2247357. 

\paragraph{Competing interests.} OpenAI provided research funding that supported participant compensation through the OpenAI Mental Health Grant. The decision to study ChatGPT was made prior to receiving funding from OpenAI. OpenAI had no role in study design, data collection, analysis, interpretation, manuscript preparation, or the decision to publish. D.Z. held internships at Microsoft and Apple in 2026. L.I. is currently employed by Google DeepMind. No competing financial interests exist related to the presented results. None of these organizations had any role in the conception, design, execution, evaluation or writing of this paper. The remaining authors declare no other competing interests.

\paragraph{Additional information.} Additional information including Supplementary Information, code, and data is available at \href{https://github.com/SALT-NLP/ai-deactivation}{the following link}.


\end{document}